\documentclass[a4paper,twoside]{article}

\usepackage{epsfig}
\usepackage{subcaption}
\usepackage{calc}
\usepackage{amssymb}
\usepackage{amstext}
\usepackage{amsmath}
\usepackage{amsthm}
\usepackage{multicol}
\usepackage{pslatex}
\usepackage{apalike}
\usepackage{algorithm2e}
\usepackage[bottom]{footmisc}
\usepackage{cite}
\usepackage{amsmath,amssymb,amsfonts}
\usepackage{algorithmic}
\usepackage[bookmarks=false]{hyperref}
\usepackage{listings}
\usepackage{graphicx}
\usepackage{textcomp}
\usepackage{xcolor}
\usepackage{tabularx, colortbl}
\usepackage{booktabs}
\usepackage{tikz}
\usepackage{makecell}
\usepackage{multirow}
\usepackage{diagbox}
\usepackage{svg}
\usepackage{gensymb}
\usepackage{subcaption}
\usepackage{adjustbox}
\usepackage[nolist]{acronym}
\usepackage{titlesec}
\usepackage{SCITEPRESS}

\usetikzlibrary{shapes,snakes}
\usetikzlibrary{calc} 
\usetikzlibrary{arrows.meta}
\usetikzlibrary{positioning}

\tikzset{
	>=Latex,
	line/.style={draw,->},
	anode/.style={rectangle,draw,
		align=center,rounded corners,minimum height=4em,font=\strut},
	bnode/.style={anode,fill=white, font=\strut},
	cnode/.style={anode, fill=cyan!20, font=\strut},
treenode/.style = {circle,
	draw=black,thick, fill=white, align=center, minimum size=1cm},
root/.style     = {treenode, font=\footnotesize},
env/.style      = {treenode, font=\footnotesize}, 
dummy/.style    = {circle,draw}
}

\newcommand*{\figref}[1]{Figure~\ref{#1}}
\def\BibTeX{{\rm B\kern-.05em{\sc i\kern-.025em b}\kern-.08em
    T\kern-.1667em\lower.7ex\hbox{E}\kern-.125emX}}

\newcommand*{\secref}[1]{Section~\ref{#1}}
\def\BibTeX{{\rm B\kern-.05em{\sc i\kern-.025em b}\kern-.08em
    T\kern-.1667em\lower.7ex\hbox{E}\kern-.125emX}}

\def\BibTeX{{\rm B\kern-.05em{\sc i\kern-.025em b}\kern-.08em
    T\kern-.1667em\lower.7ex\hbox{E}\kern-.125emX}}

\newcommand*{\lstref}[1]{Listing~\ref{#1}}
\def\BibTeX{{\rm B\kern-.05em{\sc i\kern-.025em b}\kern-.08em
    T\kern-.1667em\lower.7ex\hbox{E}\kern-.125emX}}

\lstdefinelanguage{sparql}{
morecomment=[l][\color{olivegreen}]{\#},
morestring=[b][\color{blue}]\",
morekeywords={SELECT,CONSTRUCT,DESCRIBE,ASK,WHERE,FROM,NAMED,PREFIX,BASE,OPTIONAL,FILTER,GRAPH,LIMIT,OFFSET,SERVICE,UNION,EXISTS,NOT,BINDINGS,MINUS,a},
sensitive=true
}
\definecolor{light-gray}{gray}{0.95}
\lstdefinestyle{mystyle}{
    backgroundcolor=\color{light-gray},  
    basicstyle=\ttfamily \footnotesize,
    breaklines=true,
    keywordstyle=\bfseries,
    morekeywords={SELECT, WHERE}
}
\newcolumntype{Y}{>{\centering\arraybackslash}X}

\begin{document}

\begin{acronym}
  \acro{rtm}[RTM]{\textit{Resin Transfer Molding}}
  
  \acro{odp}[ODP]{\textit{Ontology Design Pattern}}
  \acroplural{odp}[ODPs]{\textit{Ontology Design Patterns}}
  
  \acro{fe}[FE]{\textit{Finite Element}}

  \acro{parx}[ParX]{\textit{Parameter Interdependency}}

  \acro{om}[OM]{\textit{OpenMath}}

  \acro{owl}[OWL]{\textit{Web Ontology Language}}

  \acro{sa}[SA]{\textit{sensitivity analysis}}

  \acro{sparql}[SPARQL] {\textit{SPARQL Protocol and RDF Query Language}}

  \acro{sis}[SiS]{\textit{Simulation Support}}

  \acro{css}[CSS]{\textit{Capability-Skill-Service}}

  \acro{rdf}[RDF]{\textit{Resource Description Framework}}
  
\end{acronym}

\title{Semantic Capability Model for the Simulation of Manufacturing Processes}

\author{\authorname{Jonathan Reif\sup{1}\orcidAuthor{0009-0001-2079-8967}, Tom Jeleniewski\sup{1}\orcidAuthor{0009-0007-0360-4108}, Aljosha Köcher\sup{1}\orcidAuthor{0000-0002-7228-8387}, Tim Frerich\sup{2}\orcidAuthor{0000-0001-9097-712X},  Felix Gehlhoff\sup{1}\orcidAuthor{0000-0002-8383-5323} and Alexander Fay\sup{3}\orcidAuthor{0000-0002-1922-654X}}
\affiliation{\sup{1}Institute of Automation Technology, Helmut Schmidt University, Holstenhofweg 85, Hamburg, Germany}
\affiliation{\sup{2}CTC GmbH (An Airbus Company), Airbusstraße 1, Stade, Germany}
\affiliation{\sup{3}Chair of Automation, Ruhr University Bochum, Universitätsstraße 150, Bochum, Germany}
\email{\{jonathan.reif, tom.jeleniewski, aljosha.koecher, felix.gehlhoff\}@hsu-hh.de, tim.frerich@airbus.com, alexander.fay@rub.de}
}

\keywords{Simulation, Simulation Sequence, Capabilities, Ontologies, Semantic Information Model, Industry 4.0}

\abstract{Simulations offer opportunities in the examination of manufacturing processes.
They represent various aspects of the production process and the associated production systems.
However, often a single simulation does not suffice to provide a comprehensive understanding of specific process settings. 
Instead, a combination of different simulations is necessary when the outputs of one simulation serve as the input parameters for another, resulting in a sequence of simulations.
Manual planning of simulation sequences is a demanding task that requires careful evaluation of factors like time, cost, and result quality to choose the best simulation scenario for a given inquiry.\\
In this paper, an information model is introduced, which represents simulations, their capabilities to generate certain knowledge, and their respective quality criteria. 
The information model is designed to provide the foundation for automatically generating simulation sequences.
The model is implemented as an extendable and adaptable ontology.
It utilizes Ontology Design Patterns based on established industrial standards to enhance interoperability and reusability. 
To demonstrate the practicality of this information model, an application example is provided. 
This example serves to illustrate the model's capacity in a real-world context, thereby validating its utility and potential for future applications.}

\onecolumn \maketitle \normalsize \setcounter{footnote}{0} \vfill

\section{\uppercase{Introduction}}
\label{sec:introduction}
In the design phase of production processes, especially those marked by high complexity or substantial unit costs, testing various process parameter configurations with physical prototypes is both time-consuming and costly \cite{Naresh.2020}.
This issue is not limited to manufacturing companies creating entirely new production systems but also affects those aiming to improve aspects like material properties, energy efficiency, or emission levels. 
As a result, manufacturing companies are constantly redesigning processes.

Simulations emerge as a powerful tool in this context as they offer the possibility to generate process information virtually and enable cost-effective testing of different process parameter configurations within reasonable time frames \cite{Mourtzis.2020}.

However, the utility of a single simulation is often limited. 
It may not suffice to generate necessary process information. 
Instead, one simulation often relies on specific input parameters derived from preceding simulations. 
This interdependent connection between simulations necessitates an execution of simulation sequences.
In this context, the term \textit{parameter} specifically refers to input and output parameters used in a simulation corresponding to process parameters of a corresponding manufacturing process. 
It does not refer to parameters that are specific to the simulation itself, such as step size.

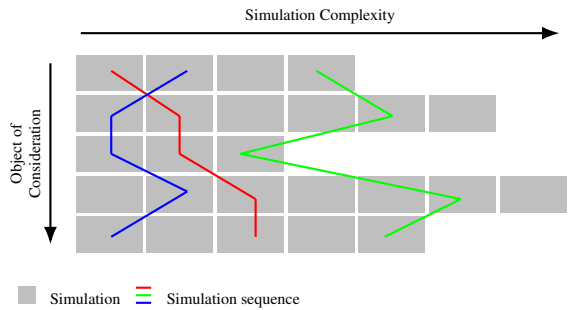
\begin{figure}[ht]
    \centering
    \renewcommand{\arraystretch}{1.2} 
    \setlength{\tabcolsep}{2pt} 
    \setlength{\arrayrulewidth}{0.5mm}
    \arrayrulecolor{white}
    \hspace{0.45cm}
    \begin{tabularx}{0.9\columnwidth}{|c|*{7}{Y|}}
        \hline
        &  &  &   &   &   &  &  \\ \hline
         & \cellcolor{gray!50} & \cellcolor{gray!50} & \cellcolor{gray!50} & \cellcolor{gray!50} &  &  & \\ \hline
          & \cellcolor{gray!50} & \cellcolor{gray!50} & \cellcolor{gray!50} & \cellcolor{gray!50} & \cellcolor{gray!50} & \cellcolor{gray!50} &  \\ \hline
          & \cellcolor{gray!50} & \cellcolor{gray!50} & \cellcolor{gray!50} &  &  &  &  \\ \hline
           & \cellcolor{gray!50} & \cellcolor{gray!50} & \cellcolor{gray!50} & \cellcolor{gray!50} & \cellcolor{gray!50} & \cellcolor{gray!50} & \cellcolor{gray!50} \\ \hline
         & \cellcolor{gray!50} & \cellcolor{gray!50} & \cellcolor{gray!50} & \cellcolor{gray!50} & \cellcolor{gray!50} &  &  \\ \hline
    \end{tabularx}
    
    \begin{tikzpicture}[overlay]
        \draw[->, thick](-2.8,3) -- (3.5,3) node [above, midway] {\tiny Simulation Complexity};
        \draw[->, thick] (-3.2,2.6) -- (-3.2,0.2) node [rotate=90, above, midway] {\tiny \begin{minipage}{1.5cm}\centering Object of Consideration\end{minipage}};
        
        \draw[red, thick] (-2.4,2.5) -- (-1.5,1.9); 
        \draw[red, thick] (-1.5,1.9) -- (-1.5,1.4); 
        \draw[red, thick] (-1.5,1.4) -- (-0.5,0.8); 
        \draw[red, thick] (-0.5,0.8) -- (-0.5,0.3);

        \draw[green, thick](0.3,2.5) -- (1.3,1.9);
        \draw[green, thick](1.3,1.9) -- (-0.7,1.4);
        \draw[green, thick](-0.7,1.4) -- (2.2,0.8);
        \draw[green, thick](2.2,0.8) -- (1.2,0.3);

        \draw[blue, thick](-1.4,2.5) -- (-2.4,1.9);
        \draw[blue, thick](-2.4,1.9) -- (-2.4,1.4);
        \draw[blue, thick](-2.4,1.4) -- (-1.4,0.9);
        \draw[blue, thick](-1.4,0.9) -- (-2.4,0.3);
    \end{tikzpicture}

    \begin{flushleft}
    \begin{tabular}{|c|c|c|c|}
        \hline
        \begin{tikzpicture}
            \node at (0,0) [draw=none, fill=gray!50, minimum width=0.2cm, minimum height=0.2cm] {};
        \end{tikzpicture} & \tiny Simulation &
        \begin{tikzpicture}
            \draw[red, thick] (0,0.2) -- (0.2,0.2);
            \draw[green, thick] (0,0.1) -- (0.2,0.1);
            \draw[blue, thick] (0,0) -- (0.2,0);
        \end{tikzpicture} & \tiny Simulation sequence \\ \hline
    \end{tabular}
    \end{flushleft}

\caption{Schematic illustration of available simulations}
\label{fig:simMatrix}
\end{figure}

In practice, these sequences are manually created by experts. 
One of the primary considerations here is to ensure the general suitability of simulations to answer specific questions. 
Another aspect relates to the quality of results and how suitable simulations can be effectively combined for this purpose.
Often simulations representing the same object of consideration, e.g., the same process step, are available in different complexities, i.e., different levels of abstraction, necessitating a selection with regard to time and resource constraints.
\figref{fig:simMatrix} illustrates this relationship. 
In this figure, various simulations (depicted as gray boxes) with increasing complexity along the X-direction are assigned to different objects of consideration along the Y-direction.
Note, that the simulations objects of consideration can be process steps, but there can also be different simulations for certain objects of consideration within a process step, e.g., different process parameters, like pressure or temperature. 
These individual simulations are interconnected through the previously described input-output relationships and can be combined to generate process information. 
This is exemplified by the colored lines. 
The figure is merely an excerpt. 
In practice, a greater number of simulations with varying levels of complexity, as well as more process steps, are available. 
This increases the complexity of selecting and generating simulation sequences.

A general assumption is that the largest part of a system should be simulated at the lowest complexity and only specific parts of interest should be simulated at a higher level of detail if necessary \cite{PuntelSchmidt.2015}. 
Due to these considerations, a manual creation of simulation sequences becomes a complex, time-consuming, and error-prone activity.
It requires a deep understanding of simulations, their parameters, and the overall production process.

In order to automatically derive a sequence of simulations, a machine-interpretable model is required. This model must describe simulations, their functionalities and relevant quality criteria, like simulation time, result quality, etc. Describing the functionalities of simulations corresponds to modeling capabilities of flexible manufacturing resources. A capability is defined as "an implementation-independent specification of a function in industrial production to achieve an effect in the physical or virtual world" \cite{Kocher&Belyaev.2023}.
Simulations can be considered to achieve an effect in the virtual world, therefore the following question arises:\\ 
\textit{How should a machine-interpretable model for planning simulation sequences be structured, building on preliminary work from the capability context and relevant standards for simulations?}

For this purpose, first specific requirements for such a model are deducted building on \cite{Reif&Jeleniewski&Fay.2023} in \secref{sec:requirements}, before drawing a comparison to related research in this field in \secref{sec:relatedwork}.
Based on this, the created information model is described and explained in detail in \secref{sec:informationmodel}, pointing out its different components.
Afterwards \secref{sec:application} describes an application example from industrial practice to show the practical potential of the created semantic model.
The paper concludes with a summary and an outlook for potential future research.

\section{\uppercase{Requirements}}
\label{sec:requirements}
As pointed out in \cite{Reif&Jeleniewski&Fay.2023} there are several requirements for an approach to automating the generation of simulation sequences:
\begin{itemize}
    \item A formal description of simulations and their parameters in a machine readable format to enable an automated selection of suitable simulations based on required process information.
    \item The consideration of quality criteria to ensure that objectives align with the capabilities of selected simulations.
    \item The consideration of parameter influences on quality criteria to ensure a useful selection of simulations.
\end{itemize}
From this emerges the necessity for a semantic model that describes simulations and their relevant properties, to facilitate the automated creation of simulation sequences \cite{Reif&Jeleniewski&Fay.2023}.
The requirements for this semantic model are explained in the following.
They can all be derived from the goal to select a simulation, respectively a sequence of simulations, that can supply a user with the right process information in a demanded quality with minimal effort.\\

\noindent \textbf{R1: Description of simulations and their capabilities in a machine readable format.}

The first requirement that the model needs to meet is the formal representation of simulations and their properties in a machine readable format. 
This format is crucial as it facilitates an automatic identification and arrangement of appropriate simulations and simulation sequences based on required process information. 
For this automation, simulations must be characterized by their input and output parameters as well as their capabilities, i.e., their ability to produce specific process information.
The potential to produce a designated output from a given input is referred to as a capability.  
Consequently, this allows for the pinpointing and selection of a fitting simulation, or a sequence of simulations, to produce desired process information through capability matchmaking, which describes the process of matching product requirements
against resource capabilities \cite{Jarvenpaa.2017}.\\

\noindent \textbf{R2: Description of selection criteria for simulations}

The second requirement involves detailing the criteria for selecting simulations. 
This is essential for choosing suitable simulations and planning sequences. 
The aim is to not only choose simulations that provide required process information for a specific inquiry but also to ensure that this information meets quality criteria and is obtained with minimal resource effort.

Thus, the criteria to be described include quality criteria to guarantee that simulations deliver information at the desired level of quality.
This also includes other selection criteria like, e.g., simulation time, including run time and lead time.  
Additionally, it is necessary to consider the influence of specific input parameters on selection criteria, such as the quality of the simulation output.
This involves documenting how input parameters affect the output to determine their influence on quality criteria.

If an input parameter has a minor impact on the overall result, less stringent quality requirements may be applied to it, and lower fidelity models may be applied to represent it.
Conversely, parameters that significantly affect the outcome should meet higher quality standards as the simulation results can only be as good as its input parameters. \cite{Reiter.2012}

By adhering to this principle, the process of generating process information can be optimized, essentially, the resource effort can be minimized, since inputs that have less impact on quality criteria can be generated using simulations that are less resource-intensive in a sequence of simulations.\\

\noindent \textbf{R3: Interoperability and reusability of the information model}

The third requirement is the interoperability and reusability of the information model. 
Models often lack the adaptability and extensibility needed for reusing them in different use cases.
For this reason, it is advisable to implement the information model in a format that offers the possibility to extend and adapt it \cite{VogelHeuser.2015}.
Furthermore, models created for specific scenarios often face challenges when applied to different situations due to varying definitions and naming conventions, making it recommendable to adhere to common semantics used in a domain \cite{Hildebrandt&Kocher.2020}. 

Additionally, to enhance acceptance in the industry, it is advantageous to utilize established industrial standards. 
Consequently, the model should be universally applicable to the considered domain and use terms that are widely accepted. 
This is achievable by conforming to industry guidelines and standards that are based on expert knowledge.

Another advantage of adhering to established standards is that in this way it is ensured that simulations of manufacturing processes are described in a consistent manner. 
This consistency allows simulations to be linked with real-world processes using the same input and output parameters, described with common semantics.  
This linkage not only enables a use of simulated process information as input for simulations but also allows for the integration of information from real processes, e.g. historical process data, or expert knowledge, like commonly used parameter values, particularly when this information meets the set quality criteria.

\section{\uppercase{Related Work}}
\label{sec:relatedwork}
In this section, which is divided into the three subsections \nameref{sec:simulationdescription},  \nameref{sec:qualitycriteriaforsimulationassessment} and \nameref{sec:descriptionofparameterinfluences}, different approaches relevant to the work presented are described and compared to the requirements mentioned in the previous section.

\subsection{Description of Simulations and their Capabilities}
\label{sec:simulationdescription}

\cite{DBLP:conf/seke/GrolingerCMS12} focus on utilizing ontologies, defined by \cite{Studer.1998Knowledgeengineering:Principles} as ”formal, explicit specification of a shared conceptualization”, for representing simulation models across various domains. Ontologies are used to provide a common understanding and facilitate operations such as querying, comparison, and inference on simulation models.
The main challenges addressed include the difficulty of extracting specific information from simulation models due to their proprietary formats and terminologies, and the complexity involved in comparing and sharing models across different simulation engines. 
The proposed solution is an ontology-based representation system that maps domain-specific simulation models to instances within a structured ontology. 
However, this method does not take into account a selection or planning of simulations or sequences of simulations, nor does it contemplate any form of simulation evaluation. 
Furthermore, it does not comply with established standards or common semantics. 

\cite{Sindelar&Novak.2016} present the development of an ontology specifically tailored for the field of modeling and simulation.
The goal is to provide a centralized, structured knowledge base that can support the interconnection, interoperability, integration, and reuse of simulation artifacts through Semantic Web Technologies.
Probable applications described include improved model discovery and sharing, enhanced capability for semantic searches, and the facilitation of knowledge transfer and reuse across different simulation studies and platforms.
However, the authors do not deal with the topics of simulation selection or planning.
Furthermore the presented ontology does not rely on industrial standards applicable for the domain.

\cite{Listl&Fischer&Weyrich.2023} present a framework for managing discrete-event simulation models through the use of knowledge graphs. 
This approach addresses the challenges associated with managing large and heterogeneous data sets in the context of production system simulations. 
The framework facilitates the integration of multiple data sources, supports model reuse, and enables various applications to leverage the structured representation of simulation models and their data. 
By adhering to the industrial standard ISA-95, the knowledge graph ensures that the data represented within the simulation models are in alignment with widely recognized manufacturing standards. 
By incorporating the industrial standard VDI 3633 \cite{VDI.2014} for simulation of logistics, material flow and production systems, the framework ensures that its approach to simulation management aligns with established best practices in the field.  
In spite of that, the approach mainly focuses on managing and integrating simulation models using structured data representations, with a significant emphasis on data management, reuse, and interoperability, without pursuing planning approaches.

Overall, various methods exist for integrating simulations into information models. 
However, most of these methods tend to not utilize industrial standards. 
Some of these approaches address the selection of simulations, though, the detailed planning of simulation sequences, particularly considering selection criteria and the influence of parameters, is not considered.

In \cite{Kocher&Belyaev.2023} a technology-independent model aimed at enhancing adaptability and flexibility in manufacturing is presented.
This so called \ac{css} model aims at describing, organizing, and executing manufacturing functions in a way that supports adaptability, reusability, and efficient management of production resources.  
The capability aspect of this \ac{css} model is a central element that bridges the gap between abstract requirements of a production process and concrete functionalities provided by physical or virtual resources.
Capabilities are abstract descriptions of functions that are required by processes and provided by resources.
According to the \ac{css} model, purely virtual functionalities may also be described as capabilities. 
However, the focus has so far been on manufacturing processes and the description of simulations using the \ac{css} model has not been considered yet.

\subsection{Quality Criteria for Simulation Assessment}
\label{sec:qualitycriteriaforsimulationassessment}
The industrial standard ISO/IEC 25010 \cite{ISO/IEC.2011} introduces a \textit{Product Quality Model} specifically designed for software, organizing software quality attributes into eight major categories, such as \textit{Functional Suitability}, \textit{Performance Efficiency}, and \textit{Compatibility}, among others. 
Each category is further detailed through various sub-characteristics. 
Additionally, other researchers have proposed comparable metrics for evaluating models and choosing simulation software, including attributes like \textit{Correctness}, \textit{Efficiency}, and \textit{Functionality} \cite{ParastooMohagheghi.2009, Fumagalli&Polenghi.2019}.
\cite{Barth.} furthermore address the critical need for a systematic evaluation of the quality of simulation models, particularly in the context of digital twins and model-based systems engineering. 
As simulation models become integral to system development and delivery, the necessity for clear quality requirements, akin to those for physical components, becomes paramount. 
Yet, such standards for simulation models are often undefined due to their novelty and the absence of universally accepted quality criteria.

\cite{Barth.} introduce a metric for objectively assessing simulation model quality, recognizing the challenge posed by the intrinsic novelty of simulations as deliverables and the existing gap in defined quality criteria. 
It emphasizes that while simulation models are assumed to possess high quality, explicit requirements or specifications are rarely articulated. 
Building on that \cite{Barth.} continue to explore the systematic evaluation of simulation model quality, focusing on expanding and clustering quality criteria.
The presented approach utilizes ISO/IEC 25010 and thus adapts and extends quality criteria from software quality domains to suit the unique aspects of simulation models, proposing a hierarchy of attributes that reflect various technical requirements and stakeholder perspectives.

\subsection{Description of Simulation Parameter Influences}
\label{sec:descriptionofparameterinfluences}
\cite{Saltelli.2008b} offer a comprehensive introduction to \ac{sa}. 
The authors provide a systematic approach to \ac{sa}, which is a method crucial for understanding how various inputs influence the outputs of models.
The authors point out how \ac{sa} can be used to gain understanding about influences of specific parameters on a simulation. 
It allows modelers to understand which parameters have the most influence on the outcome of a simulation. 
This knowledge is vital for focusing attention on the most critical parameters, thus optimizing the design and execution of further experiments or simulations.
\ac{sa} thus is a pivotal method that bridges theoretical modeling and practical application in simulations, enabling researchers and practitioners to discern how variations in input parameters influence simulation outcomes. \cite{Saltelli.2008b}

\cite{Jeleniewski&Nabizada.2023} present a semantic model developed to articulate the interdependencies between process parameters in manufacturing settings. 
This so called \ac{parx} ontology incorporates \acp{odp} derived from industrial standards, allowing for a structured and machine-readable representation of process knowledge that can be extended and reused. 
The semantic model facilitates process redesign by enabling calculations and predictions of process outputs from specific inputs.
The model incorporates the \ac{om} standard to express and manage interdependencies between process parameters within the semantic model. 
\ac{om} is a standard for representing mathematical content, enabling an exchange of mathematical expressions in a machine-readable format. An \ac{om} \ac{rdf} representation is presented in \cite{wenzel2021openmath}.
This standard is pivotal for the model's ability to detail the quantitative interdependencies and calculations associated with manufacturing processes as it is utilized to connect mathematical functions to specific process steps. \cite{Jeleniewski&Nabizada.2023}

\section{\uppercase{Simulation Ontology}}
\label{sec:informationmodel}
This section outlines the \ac{sis} information model, an ontology designed to describe simulations including their capabilities and their properties relevant for the automated planning of simulation sequences as described in \secref{sec:introduction}.
The model was designed according to the requirements outlined in \secref{sec:requirements}.

\begin{figure*}[]
  \centering
  \includegraphics[width=0.8\textwidth]{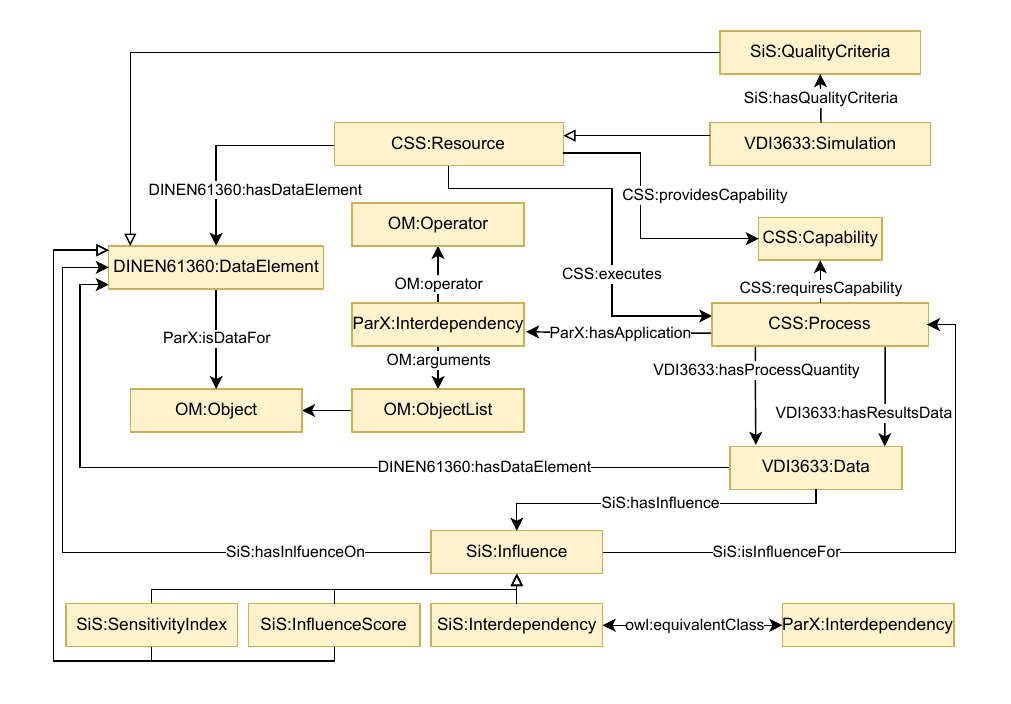}
  \caption{Alignment of the different components of the proposed semantic model for the description of simulations}
  \label{fig:simmeta}
\end{figure*}

The information model is implemented as an ontology.
This takes R1 into account, as an ontology is suited to describe simulations and their parameters in a machine readable, extensible and reusable format.
To create this model, the methodical approach from \cite{Hildebrandt&Kocher.2020} was followed. 
This approach provides a structured method for building ontologies specifically tailored for manufacturing companies. 
In this method, domain experts play a crucial role by formulating competency questions and corresponding answers. 
Competency questions hence serve as requirements that an ontology must address.
This ensures that the developed ontology considers the requirements for the automated generation of simulation sequences. 
However, it still needs to be validated afterward.
In addition, \acp{odp} are leveraged to enhance generalizability, standardization, and reduce modeling effort. 
Unlike starting from scratch, the approach of \cite{Hildebrandt&Kocher.2020} allows for high reusability of existing models, which takes into account R3. 
According to \cite{Gangemi.2009}, an \ac{odp} is a solution for recurring ontology design challenges. 
\acp{odp} are maintained separately and integrated into an alignment ontology. 
This integration enables the incorporation of additional standards and \acp{odp} in future developments, ensuring expandability. 
\acp{odp} relevant for the concept presented were identified and aligned.
The \acp{odp} used and their alignment are shown in \figref{fig:simmeta}.

As can be seen from the prefixes used in \figref{fig:simmeta} the \acp{odp} used for the \ac{sis} ontology include the \ac{css} model \cite{Kocher&Belyaev.2023}, the DIN EN 61360 (\textit{Property Descriptions of Technical Systems}) \cite{DINEN613601} and the VDI 3633 (\textit{Simulation of systems in materials handling, logistics and production}) \cite{VDI.2014} as well as the \ac{parx} ontology \cite{Jeleniewski&Nabizada.2023}, which uses the \ac{om} standard.

\newpage The information model, which is available on \textit{GitHub}~\footnote{\url{https://github.com/JonathanReif/SiS} \label{fn:sis}} can be divided into three segments: 
\begin{itemize}
    \item The description of simulations and their capabilities (R1, R3)
    \item The description of quality criteria (R2)
    \item The description of parameter influences (R2)  
\end{itemize}

In the following, these components of the overall model, as well as the industrial standards used in the information model, are described. 

The first segment focuses on the description of simulations and their capabilities.
According to \cite{Kocher&Belyaev.2023}, a \texttt{CSS:Process} is defined by its ability to convert inputs into outputs, corresponding to the fundamental process of transforming input parameters into simulation results.
A \texttt{CSS:Resource} is used for the execution of a \texttt{CSS:Process} it is connected to via the property \texttt{CSS:executes}.

In the presented approach, a simulation is categorized as a subclass of a \texttt{CSS:Resource}.
It is represented by the class \texttt{VDI3633:Simulation} \eqref{eqn:sim-Res}. 
\begin{equation}
   \scalebox{0.8}{$\displaystyle \texttt{VDI3633:Simulation} \subseteq \texttt{CSS:Resource} \label{eqn:sim-Res}$}\\
\end{equation}

In the \ac{sis} ontology, a \texttt{CSS:Process} is utilized to describe the process of generating information by simulation, which is the output derived from specific input parameters. 
As defined in the \ac{css} model, capabilities are the abstract description of a \texttt{CSS:Process}, i.e. the process of simulating information by a specific \texttt{VDI3633:Simulation}

According to the \ac{css} ontology, a resource may provide capabilities using the object property \texttt{CSS:providesCapability} and processes may require capabilities using \texttt{CSS:requiresCapability}. 
It follows from \eqref{eqn:sim-Res} that a simulation may provide a (simulation) capability, which is required by a process. 
Each process, i.e. each description of a simulation process, is further characterized using properties and classes to model input and output parameters according to VDI 3633. 
Input parameters are linked by \texttt{VDI3633:hasProcessQuantity}, output parameters by \texttt{VDI3633:hasResultsData} as VDI 3633 makes this differentiation. 
Hence, \texttt{VDI3633:hasProcessQuantity} has a domain and range defined as shown in \eqref{eqn:hasProcessQuantity_Domain} and \eqref{eqn:hasProcessQuantity_Range}, respectively.
\begin{equation}
\scalebox{0.8}{$\displaystyle \exists \; \texttt{VDI3633:hasProcessQuantity}.\top \subseteq \texttt{CSS:Process}$}
\label{eqn:hasProcessQuantity_Domain}
\end{equation}
\begin{equation}
\scalebox{0.8}{$\displaystyle \top \subseteq \forall \; \texttt{VDI3633:hasProcessQuantity}.\texttt{VDI3633:Data}$}
\label{eqn:hasProcessQuantity_Range}
\end{equation}
Domain and range of \texttt{VDI3633:hasResultsData} are defined analogously.

Both input and output parameters are modeled as \texttt{VDI3633:Data}, which is linked to a \texttt{DINEN61360:DataElement} by the object property \texttt{DINEN61360:hasDataElement}.

The DIN EN 61360 \cite{DINEN613601} standard is utilized to provide a detailed description of data elements along with their associated types and instances. The \texttt{DataElement} class is employed to characterize the properties of any data element. Each data element possesses a \texttt{DINEN61360:TypeDescription} and a \texttt{DINEN61360:InstanceDescription}.

The ontology's second segment focuses on outlining quality criteria.
R2 formulates the necessity to consider selection criteria to ensure that objectives align with capabilities of selected simulations.
Selection criteria in this context means on the one hand quality criteria that describe certain quality aspects of a respective simulation and on the other hand parameter influences. 
This means the influence of input parameters on output parameters of a simulation, respectively on different selection criteria.
A consideration of both aspects is necessary to ensure that selected simulations can produce information under consideration of quality criteria relevant for a use case. 

To set quality criteria in relation to a simulation, the object property \texttt{SiS:hasQualityCriteria} with the domain \texttt{VDI3633:Simulation} \eqref{eqn:hasQualityCriteria_Domain} and the range \texttt{SiS:QualityCriteria} \eqref{eqn:hasQualityCriteria_Range} is introduced, as can be seen in \figref{fig:simmeta}, to make it possible to assign quality criteria to the respective simulation.
The quality criteria can express different quality attributes.
\begin{equation}
\scalebox{0.8}{$\displaystyle \exists \texttt{SiS:hasQualityCriteria}.\top \subseteq \texttt{VDI3633:Simulation}$}
\label{eqn:hasQualityCriteria_Domain}
\end{equation}
\begin{equation}
\scalebox{0.8}{$\displaystyle \top \subseteq \forall \texttt{SiS:hasQualityCriteria}.\texttt{SiS:QualityCriteria}$}
\label{eqn:hasQualityCriteria_Range}
\end{equation}

As \texttt{SiS:QualityCriteria} is a subclass of \texttt{DINEN61360:DataElement}, these different attributes can be further described by a \texttt{DINEN61360:TypeDescription} elaborating on the \texttt{SiS:QualityCriteria} and by a \texttt{DINEN61360:InstanceDescription} specifying a value for the respective \texttt{SiS:QualityCriteria}.

In this way the model allows to resemble different quality criteria depending on what is deemed important for the use case considered.
For example the quality criteria for simulation models described by \cite{Barth.} could be used here to describe the quality of simulation models.

The third segment of the ontology resembles the description of parameter influences.
Influences represent the impact of an input parameter of a simulation process on the quality criteria associated with this simulation process. 
For this description the object property \texttt{SiS:hasInfluence} with the domain \texttt{VDI3633:Data} and the range \texttt{SiS:Influence} is introduced.
Additionally, the object properties \texttt{SiS:hasInfluenceOn} and \texttt{SiS:isInfluenceFor} are defined with the domain \texttt{SiS:Influence} and the range \texttt{DINEN61360:DataElement} or \texttt{CSS:Process}, respectively. 
This allows for a specification of quality criteria, respectively the simulation process that is affected by \texttt{SiS:Influence}.
An influence is expressed by the different subclasses of \texttt{SiS:Influence} shown in \figref{fig:simmeta}, see \eqref{eqn:influenceSubclasses}.
\begin{equation}
\scalebox{0.8}{$
\begin{aligned}
   &\texttt{SiS:SensitivityIndex},\\
   &\texttt{SiS:InfluenceScore}, \\
   &\texttt{SiS:Interdependency} \subseteq \texttt{SiS:Influence}
\end{aligned}$}
\label{eqn:influenceSubclasses}
\end{equation}
Several possibilities for the description of influences of input parameters on output parameters were identified.

The first possibility is the description of influences by mathematical expressions, mirroring the impact of input parameters on output parameters. 
For instance consider a composite production process, in which the glass transition temperature of resin is an important parameter.
The influence of the glas transition temperature may differ depending on whether the process temperature is in a holding phase or a heating phase. 
This variable influence can be depicted using a mathematical function. 
Those functions typically result from expert knowledge.
For a description of such influences the concept described by \cite{Jeleniewski&Nabizada.2023} was implemented in the information model.
This was realized by using the ParX ontology, which is an approach for a semantic model representing process parameters and their interdependencies \cite{Jeleniewski&Nabizada.2023}.
It is aligned with the \ac{sis} ontology through the property \texttt{ParX:hasApplication} with the domain \texttt{CSS:Process} and the range \texttt{ParX:Interdependency}.
The class \texttt{ParX:Interdependency} is equivalent to the class \texttt{SiS:Interdependency}.

The second possibility provided to express parameter influences is the option to express the influence not as a mathematical expression but as a single index.
These indices can on the one hand be assessed by the analysis of the simulation model, e.g. by a sensitivity analysis, which is a common method to determine the influence of input parameters on output parameters of a simulation \cite{Saltelli.2008b}.  
The sensitivity index that is the result of this analysis then can be expressed by the \texttt{SiS:SensitiviyIndex}. 
As \texttt{SiS:SensitivityIndex} is a subclass of \texttt{DINEN61360:DataElement} it has a \texttt{DINEN61360:TypeDescription} as well as a \texttt{DINEN61360:InstanceDescription}.
The former is used to give further information on the \texttt{SiS:SensitivityIndex} the latter to express the influence as a value in the value range $0 \leq I_n \leq 1$ with $n \in N$. 
Note, that the indices only describe a fractional contribution of one parameter on a specific quality criteria of an output $o$ which has the implication, that the sum of all parameter indices in the set $N$ for a \texttt{CSS:Process} is $1$ as described in \eqref{eq:SensitivityIndices}.

\begin{equation}\label{eq:SensitivityIndices}
   \sum_{n=1}^{N}I_{n,o} = 1 \qquad \forall o \in O
\end{equation}

If an automated analysis of a simulation is not feasible, e.g., due to the complexity of the simulation, there is the option to have the influence value evaluated by a domain expert, which can then be expressed as a \texttt{SiS:InfluenceScore}, in the same way and in the same value range as the sensitivity index.

Either way, through this description it is possible to assess whether an input parameter has a great influence on the output of a simulation and thus whether it has to be available in a certain quality or if it can be neglected.

\section{\uppercase{Application Example}}
\label{sec:application}
The model presented in this paper was developed in collaboration with an industry partner specializing in aerospace components and is evaluated within this domain.
The primary focus in the considered use case is on modeling process knowledge of the \ac{rtm} process for aircraft components during the initial design phase.
In the course of this, the possibilities of process information generation, i.e. the respective simulations, are also to be described, mapped and made usable.

In this section, the application of the presented information model through an exemplary use case is described.
The \ac{rtm} process, which is employed to manufacture fiber-reinforced plastics, consists of injecting resin under pressure into a mold that holds a fiber preform, and then proceeding with a curing phase.
The process step exemplified in the presented example is the infusion of resin into a mold.

The application example follows the method presented in \cite{Reif&Jeleniewski&Fay.2023} and demonstrates that the information model introduced in \secref{sec:informationmodel} can serve as the foundation for this method.

The method starts with identifying suitable simulations for the generation of a required process information.
In this application example, a user wants to simulate the local fill time in the process, i.e., the time until resin arrives at every point of the part.
The information model provides the possibility for a user to evaluate which simulations are capable of generating the desired information from the provided inputs.
First, the process for which the parameter \texttt{ex:LocalFillTime} is an output must be identified. 
Next, the capability required by the process needs to be determined. Finally, this required capability can be matched to the simulations that provide it.

This can be achieved via the \ac{sparql} query displayed in \lstref{lst:SPARQLSim}.
In this way this selection, normally done manually, can be supported in an automated fashion.

\begin{lstlisting}[language=sparql, keywordstyle=\color{blue}\bfseries, label={lst:SPARQLSim}, caption=A SPARQL query to identify simulations able to generate a specific output]
SELECT ?pr ?cap ?sim WHERE {
  ?pr VDI3633:hasResultsData ex:LocalFillTime .
  ?pr CSS:requiresCapability ?cap .
  ?cap a CSS:Capability .
  ?sim CSS:providesCapability ?cap .
  ?sim a VDI3633:Simulation.}
\end{lstlisting}

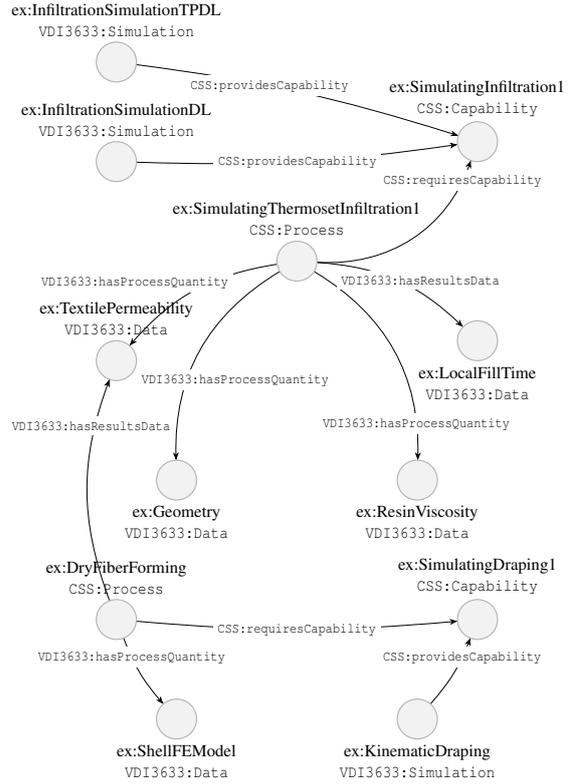
\begin{figure}
    \centering
    \resizebox{\columnwidth}{!}{
        \begin{tikzpicture}[env/.style={circle, draw=gray!60, fill=gray!10, thick, minimum size=10mm}]
            \begin{scope}[every node/.style = {font=\large, thick}, 
    			every label/.style = {font=\large, align=center}]
                \node[env, label={\text{ex:SimulatingThermosetInfiltration1} \\ \texttt{CSS:Process}}] (PO) at (0,-1) {};
                \node[env, label={\text{ex:InfiltrationSimulationDL} \\ \texttt{VDI3633:Simulation}}] (SIM) at (-4.5,1.5) {};
                \node[env, label={\text{ex:InfiltrationSimulationTPDL} \\ \texttt{VDI3633:Simulation}}] (SIMTP) at (-4.5,4) {};
                \node[env, label={\text{ex:SimulatingInfiltration1} \\ \texttt{CSS:Capability}}] (CAP) at (4.5,2) {};
                \node[env, label=below:{\text{ex:Geometry} \\ \texttt{VDI3633:Data}}] (FEM) at (-3,-6.5) {};
                \node[env, label={\text{ex:TextilePermeability} \\ \texttt{VDI3633:Data}}] (PER) at (-4.5,-3.5) {};
                \node[env, label=below:{\text{ex:ResinViscosity} \\ \texttt{VDI3633:Data}}] (VIS) at (3,-6.5) {};
                \node[env, label=below:{\text{ex:LocalFillTime} \\ \texttt{VDI3633:Data}}] (FIL) at (4.5,-3) {};
                \node[env, label={\text{ex:DryFiberForming} \\ \texttt{CSS:Process}}] (DRY) at (-4.5,-10) {};
                \node[env, label={\text{ex:SimulatingDraping1} \\ \texttt{CSS:Capability}}] (DRAP) at (4.5,-10) {};
                \node[env, label=below:{\text{ex:KinematicDraping} \\ \texttt{VDI3633:Simulation}}] (MULT) at (3,-12.5) {};
                \node[env, label=below:{\text{ex:ShellFEModel} \\ \texttt{VDI3633:Data}}] (SHELL) at (-3,-12.5) {};
            \end{scope}
            \begin{scope}[>={Stealth[black]},
				every node/.style={fill=white},
				every edge/.style={draw=black, font=\normalsize}]
				every label/.style={font=\scriptsize}
    				\path [->] (PO) edge[bend right = 37] node[pos=0.9] {\texttt{CSS:requiresCapability}} (CAP); 
    				\path [->] (SIM) edge[bend right=5] node[] {\texttt{CSS:providesCapability}} (CAP);
                    \path [->] (PO) edge[bend right=30] node[pos=0.65, xshift=1.2cm] {\texttt{VDI3633:hasProcessQuantity}} (FEM);
                    \path [->] (PO) edge[bend right=20] node[xshift=-1.5cm, yshift=0.2cm] {\texttt{VDI3633:hasProcessQuantity}} (PER);
                    \path [->] (PO) edge[bend left=30] node[pos=0.85] {\texttt{VDI3633:hasProcessQuantity}} (VIS);
                    \path [->] (PO) edge[bend left=20, right ] node[xshift=-1.5cm, yshift=0cm] {\texttt{VDI3633:hasResultsData}} (FIL);
                    \path [->] (DRY) edge[bend left=20] node[pos=0.8, xshift=-0.1cm] {\texttt{VDI3633:hasResultsData}} (PER);
                    \path [->] (DRY) edge[bend right = 5] node[] {\texttt{CSS:requiresCapability}} (DRAP);
                    \path [->] (MULT) edge[bend right=10] node[pos=0.75] {\texttt{CSS:providesCapability}} (DRAP);
                    \path [->] (SIMTP) edge[bend left=5] node[pos=0.4] {\texttt{CSS:providesCapability}} (CAP);
                    \path [->] (DRY) edge[bend right=10] node[pos=0.25] {\texttt{VDI3633:hasProcessQuantity}} (SHELL);
            \end{scope}
    
        \end{tikzpicture}
    }
    \caption{Exemplary graph excerpt resembling the capability of the simulation}
    \label{fig:SimCap}
\end{figure}

In the use case considered, two available simulations to simulate the parameter \texttt{ex:LocalFillTime} can be identified. 
These two simulations provided in the information model can be seen in \figref{fig:SimCap}, an infiltration simulation based on Darcy’s law, which describes the flow of a fluid through a porous medium and an infiltration simulation based on two phase Darcy's law, which describes immiscible two-phase flow in porous media.

As can be seen in \figref{fig:SimCap}, the simulations are expressed as \texttt{VDI3633:Simulation}.
As both simulations for the thermoset infiltration can generate the same output from the same input parameters they are both connected to the same capability \texttt{ex:SimulatingInfiltration1} via the property \texttt{CSS:providesCapability}.
The simulation process generating \texttt{ex:LocalFillTime} out of the respective inputs is expressed as \texttt{ex:SimulatingThermosetInfiltration1}.
As the respective object properties express whether \texttt{VDI3633:Data} is an input or an output of the process the instances \texttt{ex:Geometry}, \texttt{ex:TextilePermeability} and \texttt{ex:ResinViscosity} are connected to \texttt{ex:SimulatingThermosetInfiltration1} via \texttt{VDI3633:hasProcessQuantity} and \texttt{ex:LocalFilltime} is connected via \texttt{VDI3633:hasResultsData}.

The next step is the selection of a suitable simulation based on use case specific quality criteria. 
This quality criteria can be queried via \ac{sparql} as well as can be seen in \lstref{lst:SPARQLQual}.
In \lstref{lst:SPARQLQual} only the quality criteria for \texttt{ex:InfiltrationSimulationDL} is queried, the quality criteria for other simulations can be queried accordingly.

\begin{lstlisting}[
  language=sparql, keywordstyle=\color{blue}\bfseries, label={lst:SPARQLQual}, caption=A SPARQL query to get quality criteria of a simulation]
SELECT ?qual ?id ?val
WHERE {ex:InfiltrationSimulationDL a VDI3633:Simulation .
  ?sim SiS:hasQualityCriteria ?qual .
  ?qual DINEN61360:has_Instance_Description ?id .
  ?id DINEN61360:value ?val .}
\end{lstlisting}

\begin{figure}
    \centering
    \resizebox{\columnwidth}{!}{
        \begin{tikzpicture}[env/.style={circle, draw=gray!60, fill=gray!10, thick, minimum size=10mm}]
            \begin{scope}[every node/.style = {font=\large, thick}, 
    			every label/.style = {font=\large, align=center}]
                \node[env, label={\text{ex:InfiltrationSimulationDL} \\ \texttt{VDI3633:Simulation}}] (SIM) at (3,5.5) {};
                \node[env, label={\text{ex:ResultAccuracyInfSimDL} \\ \texttt{SiS:QualityCriteria}}] (QUAL) at (-5,5.5) {};
                \node[env, label=below:{\text{ex:ResultAccuracyInfSimDL\_ID} \\ \texttt{DINEN61360:InstanceDescription}}] (IDQ) at (-5,3) {};
                \node[env, label={\text{ex:ResultAccuracy} \\ \texttt{DINEN61360:TypeDescription}}] (TDQ) at (3,2) {};
                \node[] (idqual) at (-0.2,3) {$0.7$};
                \node[env, label={\text{ex:ResultAccuracyInfSimTPDL} \\ \texttt{SiS:QualityCriteria}}] (QUALTP) at (-1,-1) {};
                \node[env, label=below:{\text{ex:ResultAccuracyInfSimTPDL\_ID} \\ \texttt{DINEN61360:InstanceDescription}}] (IDQTP) at (3,-4) {};
                \node[] (idqualTP) at (3,-1) {$0.8$};
                \node[env, label=below:{\text{ex:InfiltrationSimulationTPDL} \\ \texttt{VDI3633:Simulation}}] (SIMTP) at (-5,-4) {};
            \end{scope}
            \begin{scope}[>={Stealth[black]},
				every node/.style={fill=white},
				every edge/.style={draw=black, font=\normalsize}]
				every label/.style={font=\scriptsize};
    				\path [->] (SIM) edge[bend right = 5] node[] {\texttt{SiS:hasQualityCriteria}} (QUAL); 
    				\path [->] (QUAL) edge[] node[pos=0.3] {\texttt{DINEN61360:hasInstanceDescription}} (IDQ);
                    \path [->] (QUAL) edge[bend left=20] node[pos=0.7] {\texttt{DINEN61360:hasTypeDescription}} (TDQ);
                    \path [->] (IDQ) edge[] node[] {\texttt{DINEN61360:value}} (idqual);
                    \path [->] (QUALTP) edge[bend right=35] node[pos=0.85] {\texttt{DINEN61360:hasTypeDescription}} (TDQ);
                    \path [->] (QUALTP) edge[bend right=20] node[pos=0.4, xshift=-1.0cm] {\texttt{DINEN61360:hasInstanceDescription}} (IDQTP);
                    \path [->] (SIMTP) edge[bend left=20] node[pos=0.5] {\texttt{SiS:hasQualityCriteria}} (QUALTP);
                    \path [->] (IDQTP) edge[] node[pos=0.7] {\texttt{DINEN61360:value}} (idqualTP);
            \end{scope}
        \end{tikzpicture}
    }
    \caption{Exemplary graph excerpt resembling the quality criteria of the simulation}
    \label{fig:SimQual}
\end{figure}
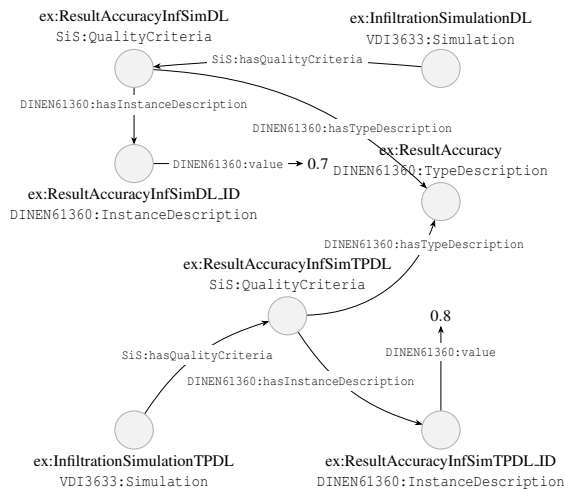

For the selection of a simulation the quality criteria are significant.   
The quality criteria relevant to this use case include lead time for preparing the simulation, simulation time, technological maturity of the simulation, and the result’s accuracy.
For simplification purposes the exemplary graph in \figref{fig:SimQual} only shows the quality criteria result accuracy.
It is expressed as \texttt{ex:ResultAccuracyInfSimDL} for the simulation \texttt{InfiltrationSimulationDL} and as \texttt{ex:ResultAccuracyInfSimTPDL} for the simulation \texttt{InfiltrationSimulationTPDL} as both simulations have a different accuracy.
Accuracy in this case means the consideration of phenomena such as pore formation, which can be simulated with \texttt{ex:InfiltrationSimulationTPDL}.
Furthermore, both instances of \texttt{SiS:QualityCriteria} have a \texttt{DINEN61360:InstanceDescription}, which makes it possible to express a value for the accuracy.
As can be seen in \figref{fig:SimQual}, the returned value for the quality criterion "Result Accuracy" for \texttt{ex:InfiltrationSimulationDL} would be $0.7$ in this example. The value for \texttt{ex:InfiltrationSimulationTPDL} would be $0.8$, indicating a higher result accuracy for the latter.

This helps users to make an informed decision when choosing the appropriate simulation for their needs.
For this example the user chooses the \texttt{ex:InfiltrationSimulationDL} because a higher result accuracy is not needed for this use case.
If the consideration of pore formation is relevant a higher result accuracy should be chosen and the simulation \texttt{ex:InfiltrationSimulationTPDL} would be better suited, which is information, that can be provided in the \texttt{DINEN61360:TypeDescription} of the quality criteria \texttt{es:ResultAccuracy}.
The trade-off here is the additional simulation time this simulation requires, which is also modeled as \texttt{SiS:QualityCriteria} but is omitted, for clarity reasons, in \figref{fig:SimQual}.

The next step in the method is the identification of necessary inputs for the simulation process. 
This can be done via the \ac{sparql} query displayed in \lstref{lst:SPARQLInp}
\begin{lstlisting}[language=sparql, keywordstyle=\color{blue}\bfseries, label={lst:SPARQLInp}, caption=A SPARQL query to get inputs of a simulation]
SELECT ?cap ?pr ?pq WHERE {
  ex:InfiltrationSimulationDL CSS:providesCapability ?cap .
  ?pr CSS:requiresCapability ?cap .
  ?pr VDI3633:hasProcessQuantity ?pq .}
\end{lstlisting}

In the presented example, the simulation uses the geometry of the textile (i.e., the composite material) discretized by a polygon mesh, the textile permeability, and the resin viscosity as inputs as can be seen in \figref{fig:SimCap}.

The next step in the method is to determine if the identified input parameters are known to the user or if they must be generated via simulation.
For this example the user knows the necessary input parameters except for the textile permeability, which the user has to simulate then.
In this way the cycle starts again and the user has to identify a suitable simulation again as shown in \lstref{lst:SPARQLSim}.
The permeability of the fiber is dictated by the draping of said fiber. 
Therefore a draping simulation can be used to determine the permeability as opposed to measuring it experimentally. 
The model takes this into account by representing the required input parameter \texttt{ex:TextilePermeability} of the simulation process \texttt{ex:SimulatingThermosetInfiltration1} as the output parameter of the simulation process \texttt{ex:DryFiberForming} which requires the capability \texttt{ex:SimulatingDraping1} offered by the simulation \texttt{ex:KinematicDraping}.
This simulation in turn requires a Shell-FE-Model as an input.
In this way it is possible to express sequences of simulations resulting from this relation between input and output parameters of connected simulations.
The output \texttt{ex:LocalFillTime} of the simulation process \texttt{ex:SimulatingThermosetInfiltration1} can in turn be used as further input for succeeding simulations like, e.g., a distortion prediction with a thermochemical simulation, which is not pictured in \figref{fig:SimCap} for clarity reasons.
The user applies the presented method again taking into account the quality criteria important for him.
Another relevant aspect in this context is the influence that an input parameter has on the important quality criteria. 
If the influence is high there should be more consideration in choosing the right simulation to generate required input parameters than if the influence is low.
The influence of the input parameters for a specific simulation can be queried as shown in \lstref{lst:SPARQLInf}.

\begin{lstlisting}[
  language=sparql, keywordstyle=\color{blue}\bfseries, label={lst:SPARQLInf}, caption=A query formulated in SPARQL to identify parameter influences]
SELECT ?cap ?pr ?inf ?inp ?id ?val
WHERE {
  ex:InfiltrationSimulationDL CSS:providesCapability ?cap.
  ?pr CSS:requiresCapability ?cap.
  ?inf SiS:isInfluenceFor ?pr.
  ?inp SiS:hasInfluence ?inf.
  ?inf DINEN61360:has_Instance_Description ?id.
  ?id DINEN61360:value ?val.}
\end{lstlisting}

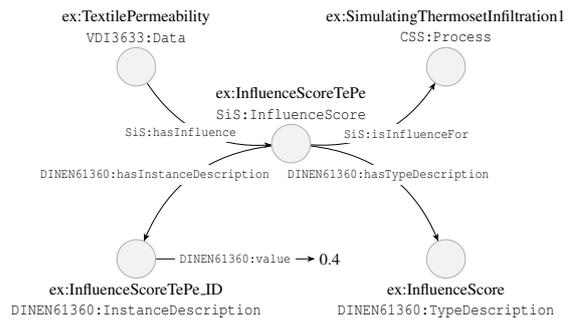
\begin{figure}
    \centering
    \resizebox{\columnwidth}{!}{
        \begin{tikzpicture}[env/.style={circle, draw=gray!60, fill=gray!10, thick, minimum size=10mm}]
            \begin{scope}[every node/.style = {font=\large, thick}, 
    			every label/.style = {font=\large, align=center}]
                \node[env, label={\text{ex:TextilePermeability} \\ \texttt{VDI3633:Data}}] (PER) at (-4,4) {};
                \node[env, label={\text{ex:InfluenceScoreTePe} \\ \texttt{SiS:InfluenceScore}}] (INF) at (0, 2) {};
                \node[env, label={\text{ex:SimulatingThermosetInfiltration1} \\ \texttt{CSS:Process}}] (PO) at (4,4) {};
                \node[env, label=below:{\text{ex:InfluenceScoreTePe\_ID} \\ \texttt{DINEN61360:InstanceDescription}}] (IDI) at (-4,-1) {};
                \node[env, label=below:{\text{ex:InfluenceScore} \\ \texttt{DINEN61360:TypeDescription}}] (TDI) at (4,-1) {};
                \node[] (val) at (1,-1) {$0.4$};
            \end{scope}
            \begin{scope}[>={Stealth[black]},
				every node/.style={fill=white},
				every edge/.style={draw=black, font=\normalsize}]
				every label/.style={font=\scriptsize}
    			    \path [->] (PER) edge[bend right =30] node[xshift=-0.5cm] {\texttt{SiS:hasInfluence}} (INF); 
                    \path [->] (INF) edge[bend right = 30] node[pos=0.3, xshift=1.3cm, yshift=0.2cm] {\texttt{SiS:isInfluenceFor}} (PO);
                    \path [->] (INF) edge[bend right = 30] node[xshift=-1cm] {\texttt{DINEN61360:hasInstanceDescription}} (IDI);
                    \path [->] (INF) edge[bend left = 30] node[] {\texttt{DINEN61360:hasTypeDescription}} (TDI);
                    \path [->] (IDI) edge[] node[] {\texttt{DINEN61360:value}} (val);
            \end{scope}
    
        \end{tikzpicture}
    }
    \caption{Exemplary graph excerpt resembling the parameter influences of the simulation}
    \label{fig:SimInf}
\end{figure}

In this application example, the influence of the input parameters on the output parameters was assessed by a domain expert. 
It was determined that textile permeability and quality of the geometry discretized by a polygon mesh have a high influence on the simulation output and therefore get a score of $0.4$ each, while resin viscosity has a medium influence and therefore gets a score of $0.2$. 
In \figref{fig:SimInf} only the influence of textile permeability for \texttt{SimulatingThermosetInfiltration1} is pictured for simplification purposes.
Other influences are modeled accordingly.

The result of this method is a simulation sequence in which certain simulations have to be executed before others to deliver necessary input parameters for the former.
In this example, the simulation \texttt{ex:KinematicDraping} must be executed before the simulation \texttt{ex:InfiltrationSimulationDL} or \texttt{ex:InfiltrationSimulationTPDL}, as shown in \figref{fig:SimCap}.

The information model presented allows representing the connection between different simulations by modeling both input and output parameters of those simulations as \texttt{VDI3633:Data}, facilitating the description of simulation sequences.
Describing quality criteria as shown in \figref{fig:SimQual} not only acknowledges the overall ability of simulations to produce specific knowledge, but also allows for the consideration of these quality criteria when determining the most suitable simulation for a particular scenario.

The structured description of quality criteria supports the consideration of such criteria in simulation selection and planning, a complex process done by experts.
By describing it in a machine readable format this consideration can be supported in an automated fashion as well.
Another aspect the model supports is the evaluation and comparison of different planned simulation sequences based on relevant quality criteria.

Evaluating and describing the impact of specific parameters allows for the planning of efficient simulation sequences true to the motto \textit{as precise as necessary but as imprecise as possible}.
In this example result accuracy is of high importance.
For this reason it can be derived that \texttt{TextilePermeability}, having a high influence, should be taken into particular consideration, which in turn sets quality requirements for the simulation \texttt{ex:KinematicDraping} that generates this input parameter.

The application of the \ac{sis} ontology to the example described in this section was carried out in collaboration with a simulation expert from the manufacturing domain. 
This collaboration confirmed the assumption that the ontology is suitable for describing simulations for manufacturing processes, considering important aspects for selection and planning. 
Consequently, the ontology supports the selection and planning of simulations by structuring necessary information and making it accessible to users.

The modeled example including the mentioned instances not pictured in the exemplary graph excerpts can be found on GitHub\footref{fn:sis}.

\section{\uppercase{Conclusion and Future Work}}
\label{sec:conclusion}
In conclusion, this paper has presented a novel information model designed to address the challenges inherent in the planning of simulation sequences, according to the question raised at the beginning.
The model represents simulations, their ability to generate specific knowledge, and their respective quality criteria, thereby facilitating the generation of simulation sequences.
The development of the model was guided by specific requirements and has proven practical in an application example. 
This underscores the model's utility in providing necessary information to support users in simulation selection and planning.
However, instantiating the information model requires expertise in both the semantic web domain and the simulation domain, as well as a solid understanding of the proposed information model.
Furthermore, the model on its own only offers limited added value to a user, as it is intended to serve as a foundation for further developments and applications that build upon it.
In this way the information model presented in this paper paves the way for further research into automated support for simulation sequence creation, with the ultimate goal of reducing the effort required to generate necessary process information.

In terms of future work, there are several potential extensions and developments building on the work presented in this paper.
The first aspect is the integration of the presented information model into a system that provides an intuitive user interface. 
This allows users to interact with the information model in a way that is significantly useful to them, as the formulation of \ac{sparql} queries is not feasible for most users.
In this context, the question of how the information model can be automatically instantiated for existing simulations and during simulation creation could be examined more closely.
The second important area of development is the creation of a planning logic for automated selection of simulations and planning of simulation sequences, as users would benefit significantly from a higher degree of automation in simulation planning.
In this scenario, it could be beneficial to examine the evaluation of quality criteria for specific use cases more closely, as well as the uncertainties that are inherently associated with the assessment of this quality criteria.
Developing a prototype for the automated generation of simulation sequences is a practical step towards implementing this planning logic.
As a possible extension to the information model it could be beneficial to incorporate a skill component, meaning the executable implementation of the abstract functions in addition to the capabilities describing these functions. \cite{Kocher&Belyaev.2023} 
This would allow for a representation of how to execute simulations.  
In this context, it would be valuable to assess the feasibility of automating the execution of simulations, which would involve evaluating the extent to which simulation parameterization and execution can be automated, identifying limitations or challenges that may arise.

\section*{\uppercase{Acknowledgements}}

This contribution originates from the \textit{LaiLa} project, funded by \textit{dtec.bw – Digitalization and Technology Research Center of the Bundeswehr} which we gratefully acknowledge. \textit{dtec.bw} is funded by the \textit{European Union – NextGenerationEU}

\bibliographystyle{apalike}
{\small
\bibliography{example}}

\end{document}